\documentclass[aps,prb,preprint,a4paper,showpacs]{revtex4-1}
\usepackage{graphicx}
\usepackage{amsmath}
\usepackage{hyperref}
\usepackage[svgnames]{xcolor}

\begin{document}
\title{Low-energy magnetoelectric control of domain states in exchange-coupled heterostructures} 
\author{Muftah Al-Mahdawi}
\email{mahdawi@ecei.tohoku.ac.jp}
\author{Satya Prakash Pati}
\author{Yohei Shiokawa}
\author{Shujun Ye}
\author{Tomohiro Nozaki}
\author{Masashi Sahashi}
\affiliation{Department of Electronic Engineering, Tohoku University, Sendai 980-8579, Japan}
\date{\today}

\newcommand{\cro}{Cr$_2$O$_3$~}

\begin{abstract}
The electric manipulation of antiferromagnets has become an area of great interest recently for zero-stray-field spintronic devices, and for their rich spin dynamics. Generally, the application of antiferromagnetic media for information memories and storage requires a heterostructure with a ferromagnetic layer for readout through the exchange-bias field. In magnetoelectric and multiferroic antiferromagnets, the exchange coupling exerts an additional impediment (energy barrier) to magnetization reversal by the applied magnetoelectric energy. We proposed and verified a method to overcome this barrier. We controlled the energy required for switching the magnetic domains in magnetoelectric \cro films by compensating the exchange-coupling energy from the ferromagnetic layer with the Zeeman energy of a small volumetric spontaneous magnetization found for the sputtered \cro films. Based on a simplified phenomenological model of the field-cooling process, the magnetic and electric fields required for switching could be tuned. As an example, the switching of antiferromagnetic domains around a zero-threshold electric field was demonstrated at a magnetic field of 2.6 kOe.
\end{abstract}
\maketitle
\section{Introduction}
Control of magnetic states by an electrical voltage has garnered a mainstream status in magnetism and spintronics research \cite{ohno_2000,chiba_2003,borisov_2005,weisheit_2007,he_2010,endo_2010,chiba_2011,shiota_2011,matsukura_2015}. Even though those efforts shed more light on the physics of magnetism, the main driving trend is to realize ultra-low-energy technologies for information storage and processing. The most promising techniques rely on voltage control of magnetic anisotropy and coercivity \cite{ohno_2000,chiba_2003,weisheit_2007,endo_2010,chiba_2011,shiota_2011}. On the other hand, electric control of magnetic states can also be realized based on magnetoelectric (ME) or multiferroic media \cite{fiebig_2005-1,eerenstein_2006}. Their spin configurations are not symmetric with regard to the directions of electric and magnetic fields, \emph{i.e.,}~they lack space inversion and time reversal symmetries \cite{dzyaloshinskii_1960}. The lack of such symmetries results in a coupling of induced (spontaneous) magnetic and electric polarizations in ME (multiferroic) media \cite{eerenstein_2006,iyama_2013}. In general, ME and multiferroic materials are relatively few \cite{hill_2000}, and many of them have low transition temperatures. Utilizing the most promising materials requires a heterostructure, where electric control of the exchange bias on a proximate ferromagnetic (FM) layer is used for readout \cite{borisov_2005,he_2010,wu_2013,heron_2014}.

Chromium oxide (Cr$_2$O$_3$) was the first material that demonstrated a linear ME effect \cite{dzyaloshinskii_1960,astrov_1960,odell_1970-1,fiebig_2005-1}. Cr$_2$O$_3$ is a corundum-type collinear antiferromagnet (AFM) with a N\'eel temperature $T_N$ = 307 K, slightly higher than the room temperature. Cr$_2$O$_3$ has two possible spin configurations aligned along the \emph{c}-axis of the rhombohedral unit cell, either $L^+$ ($\uparrow$ $\downarrow$ $\uparrow$ $\downarrow$) or $L^-$ ($\downarrow$ $\uparrow$ $\downarrow$ $\uparrow$). The spin configuration is not symmetric under either space inversion or time reversal operations, and a magnetization $M$ (electric polarization $P$) is induced by an applied electric $E$ (magnetic $H$) field. The induced coupling of the electric and magnetic orders is represented by a linear ME susceptibility $\alpha=dM/dE$ ($\alpha=dP/dH$). ME control of the exchange bias on an FM layer was first demonstrated in single-crystal slabs \cite{borisov_2005,he_2010}. The recent progress in fabricating high-quality films of Cr$_2$O$_3$ by sputtering \cite{shiratsuchi_2011,pati_2015,shimomura_2016,[][{. accepted in \emph{Physical Review B}.}]pati_2016} has paved the way for ME-controlled devices \cite{ashida_2014,ashida_2015,toyoki_2015,toyoki_2015-1,belashchenko_2016}. The ME properties of Cr$_2$O$_3$ films were shown to be close to those of the bulk crystals \cite{borisov_2016}. However, the required ME energy for switching magnetic domains in films was 3--4 orders of magnitude higher compared to that in bulk single crystals. It can be deduced from the model presented by Borisov \emph{et al.}~\cite{borisov_2005} that the interfacial exchange coupling with the FM layer exerts an additional energy barrier to be overcome by the ME energy \cite{shiratsuchi_2016}. This barrier is inversely proportional to the thickness of the AFM layer \cite{toyoki_2014,ashida_2014,shiratsuchi_2016}. A dilemma ensues; a small ME switching energy in thin films (20--50 nm) while having a large exchange-bias is required for reliable device operation.

One of the main features of AFMs is the lack of a net stray magnetization. However, a weak magnetization from Cr$_2$O$_3$ films has been observed and they were attributed to a roughness-invariant boundary magnetization at the surface \cite{he_2010, belashchenko_2010,wu_2011,fallarino_2015}. Contrarily, we show in this work that a small spontaneous magnetization $M_0$ originates within the volume of the films from the spin ordering of \cro. We utilized the Zeeman energy of $M_0$ to compensate the energy from an exchange-coupled FM layer, and achieved a switching threshold of \cro domains at a finite magnetic field and a zero electric field.

\section{Magnetoelectric field-cooling process}
We start by describing a simplified phenomenological model of the switching of AFM domains during magnetoelectric field-cooling (MEFC) of Cr$_2$O$_3$ films under magnetic field $H_\mathrm{fr}$ and electric field $E_\mathrm{fr}$. We base the model on Ref.~\cite{borisov_2005}, by including the effect of $M_0$ and accounting for thermal fluctuations at $T_N$. The Cr spins in $L^+$ ($L^-$) domain point inward (outward) the shared triangle between the oxygen octahedra [the grayed triangle in Fig.\ref{fig:schem}(a)] after a $+$MEFC ($-$MEFC) \cite{brown_1998}. We define the AFM staggered magnetization vector $l$ as parallel to the sublattice spin beneath the shared oxygen triangle. Hence, the angle between $l$ and $+z$ direction $\theta$ is $0$ ($\pi$) for $L^+$ ($L^-$) domain state [Fig.~\ref{fig:schem}(a)]. The dominance of either $L^\pm$ domain is determined by the energies affecting the AFM layer at $T_N$ \cite{borisov_2005}. The relevant are the ME energy $W_\mathrm{ME}$, the Zeeman energy of Cr$_2$O$_3$ magnetization $W_\mathrm{ZM}$, and the exchange coupling to the FM layer $W_\mathrm{EX}$. We consider the film to be an ensemble of uncorrelated particles at $T_N$, and each of them contains a single uniform classical spin within a short-range-order volume $V$ set by the grain size \cite{[][{. Submitted to \emph{Journal of Physics D}.}]al-mahdawi_2016}. Then the total free energy in cgs units for a single particle is

\begin{align}\label{eq:freeE}
W/V &= W_\mathrm{ME} + W_\mathrm{ZM} + W_\mathrm{EX} \nonumber\\
 &= -\alpha E_\mathrm{fr} H_\mathrm{fr} \cos\theta - M_0 H_\mathrm{fr}\cos\theta + \frac{J_K}{t} \cos\theta \nonumber \\
 &\equiv W_0 \cos\theta\,,
\end{align}
where $M_0$, $t$, and $J_K$ are the magnetization per unit volume, the thickness of the AFM layer, and the exchange coupling energy between the FM and AFM layers at the interface. For simplicity, $\theta$ is assumed to be uniform along the $z$-direction. This simplification is more valid for high $H_\mathrm{fr}$, as $W_\mathrm{ZM}$ and $W_\mathrm{ME}$ will be much larger than $W_\mathrm{EX}$. The macroscopic values of $M_0$, $J_K$, and $\alpha$ at $T_N$ are zero. However, this is due to the thermal averaging of the order parameter \cite{rado_1961,mostovoy_2010}. Equation \ref{eq:freeE} describes the energy of single-domain particles smaller than the short-range-order length at $T_N$. Therefore, the effective values of $M_0$ and $J_K$ in Eq.~\ref{eq:freeE} should be close to the measurable values at a low-temperature limit. On the other hand, due to the larger thermal fluctuations at $T_N$, the exchange-driven $\alpha (T_N)$ on a single-particle scale should be larger than the macroscopic peak value at 240--270 K. The choice of signs in Eq.~\ref{eq:freeE} is such that the sign of $W_\mathrm{ME}$ produces an $L^+$ ($L^-$) domain for a positive (negative) product of $E_\mathrm{fr}$ and $H_\mathrm{fr}$. The signs of $W_\mathrm{ZM}$ and $W_\mathrm{EX}$ were chosen to prefer $l$ parallel and antiparallel to $H_\mathrm{fr}$, respectively [Fig.~\ref{fig:schem}(b)]. $J_K$ is defined as positive for an antiparallel coupling between FM's magnetization and $l$. The reasoning for these choices is based on the experiments and discussions presented latter.

During the transition from the fluctuating state to the ordered state at $T_N$ during cooling, the AFM domains are stabilized within a few kelvins below $T_N$ by an increase of the crystalline anisotropy barrier \cite{al-mahdawi_2016}. The condition for the preferable domain is $\partial^2 W / \partial \theta^2 > 0$. An $L^+$ ($L^-$) domain is more stable when $W_0 < 0$ ($W_0 > 0$). At the switching threshold condition of $W_0 = 0$, an equal number of $L^\pm$ domains are present. For a fixed $H_\mathrm{fr}$, the threshold electric field for switching $E_\mathrm{th}$ is . 

\begin{align}
E_\mathrm{th}&=-\frac{M_0}{\alpha} + \frac{J_K}{\alpha t}\frac{1}{H_\mathrm{fr}}\nonumber\\
&\equiv E_M + \frac{\mathit{EH}_J}{\lvert H_\mathrm{fr} \rvert}\,.\label{eq:Eth}
\end{align}
The absolute value of $H_\mathrm{fr}$ in Eq.~\ref{eq:Eth} is to account for the switching of the FM spin direction by a negative $H_\mathrm{fr}$, and hence a change in the sign of $W_\mathrm{EX}$.

The $E_\mathrm{th}$--$1/H_\mathrm{fr}$ plot is a line with a slope determined by the strength of $J_K/t$ \cite{borisov_2005,toyoki_2015}. For thinner Cr$_2$O$_3$ films and a stronger exchange bias of the FM layer, a higher electric field is required compared to thick slabs with a weak exchange bias \cite{borisov_2005,ashida_2014,ashida_2015,toyoki_2015,toyoki_2015-1}. However, with the presence of $M_0 \parallel l$, a negative \emph{y}-intercept is found. Therefore, a zero $E_\mathrm{th}$ can be realized at a finite magnetic field $H_{E0}$. At $H_{E0}$, the Zeeman energy of $M_0$ balances the exchange coupling energy $J_K/t$. Then a small electric field can be used to choose either of the AFM domain states, and the dilemma between having a higher $J_K/t$ ratio and lower electric and magnetic fields is resolved.

The probability of switching after MEFC is found from the averaged domain state $\langle L \rangle$

\begin{equation}
\langle L \rangle \equiv \frac{v^+ - v^-}{v^+ + v^-}\,,
\end{equation}
where $v^\pm$ are the volume proportions of $L^\pm$ domains. The probability distribution $P_\Theta$ of $\theta$ for each particle at $T_N$ is given by

\begin{equation}
P_\Theta(\theta) = \frac{W_0 V}{\sinh(W_0 V)} \sin\theta \exp\left(\frac{-W_0 V\cos\theta}{k_B T_N}\right)\,,
\end{equation}
where $k_B$ is the Boltzmann constant, and $W_0 V / \sinh(W_0 V)$ is the normalization factor. As the AFM domains are forced into either domain state by the anisotropy barrier, the switching probability $\langle L \rangle$ after MEFC can be found as the expectation value of $g\left(\theta\right) = \mathrm{sgn}\left(\theta-\pi/2\right)$, where $\mathrm{sgn}$ is the signum function, as follows:

\begin{align}\label{eq:Ftanh}
\langle L \rangle =& \int_0^{\pi} g P_\Theta d\theta = \tanh \left(\frac{-W_0 V}{2k_B T_N}\right)  \nonumber \\ 
                  \equiv &\tanh \left(\frac{E_\mathrm{fr} - E_\mathrm{th} } {\lambda_E}\right)\,, 
\end{align}
where the finite probability of switching around $E_\mathrm{th}$ is described by $\lambda_E$:

\begin{equation}\label{eq:lE}
\lambda_E = \frac{2 k_B T_N}{\alpha V} \frac{1}{H_\mathrm{fr}} \equiv  \frac{\mathit{EH}_T}{H_\mathrm{fr}}\,.
\end{equation}

\section{Experimental verification}
We compared the presented model with experimental observations made on deposited films and a bulk slab of \cro. The description of samples preparation, structural characterizations, and measurement setup were reported elsewhere \cite{shimomura_2016,borisov_2016,al-mahdawi_2016}. Sputter-deposited films of Pt (25)/Cr$_2$O$_3$ (500)/top layer were prepared over \emph{c}-Al$_2$O$_3$ substrates, where the numbers in parentheses are the calibrated thicknesses in nanometers. Two types of films were prepared. In the film sample woEB (without exchange-bias), a single Pt (25) top layer was deposited. In the film sample wEB (with exchange-bias), the top layer was an FM exchange-coupled through a metal spacer, with the composition of Pt (1.1)/Co (1)/Pt (25). The magnetic and magnetoelectric properties were measured by a magnetometer based on a superconducting quantum interference device. The average domain state was determined from measurements of peak $\alpha$ in the setup described in \cite{al-mahdawi_2016}. The parallel ME susceptibility $\alpha$ was measured by applying an out-of-plane \emph{ac} electric field and measuring the in-phase component of the electrically induced magnetization of the Cr$_2$O$_3$ films \cite{borisov_2007,borisov_2016}. In contrast to measuring the exchange bias on the FM layer, the present scheme has the benefit of probing the AFM domain state inside the film, regardless of the presence of an FM layer [See appendix \ref{sec:appx_alpha}]. The low-temperature limit of exchange-bias in the sample wEB was 570 Oe, corresponding to $J_K = 8 \times 10^{-3} \mathrm{erg/cm}^2$ [See appendix \ref{sec:appx_Hex}].

\begin{figure}
	\caption{(a) The magnetization areal density of \cro films shows a linear dependence on thickness, with a negligible surface component. The dashed area is the expectation from a thickness-independent boundary magnetization. The magnetization measurements from~\cite{fallarino_2015} are also plotted. (b, c) The temperature dependence of (b) the thermoremnant magnetization (TRM), and (c) the magnetoelectric susceptibility $\alpha$ of the sample without an exchange-bias (woEB) after cooling in a positive and a negative $H_\mathrm{fr}$. The inset in (c) shows coinciding transition temperatures from TRM and $\alpha$. (d) The temperature dependence of $\alpha$, for the sample wEB. (c,d) For the same direction of $H_\mathrm{fr}$, the opposite \cro domains are generated in the samples woEB and wEB.}
	\label{fig:coupling}
	\includegraphics[width=0.6\textwidth]{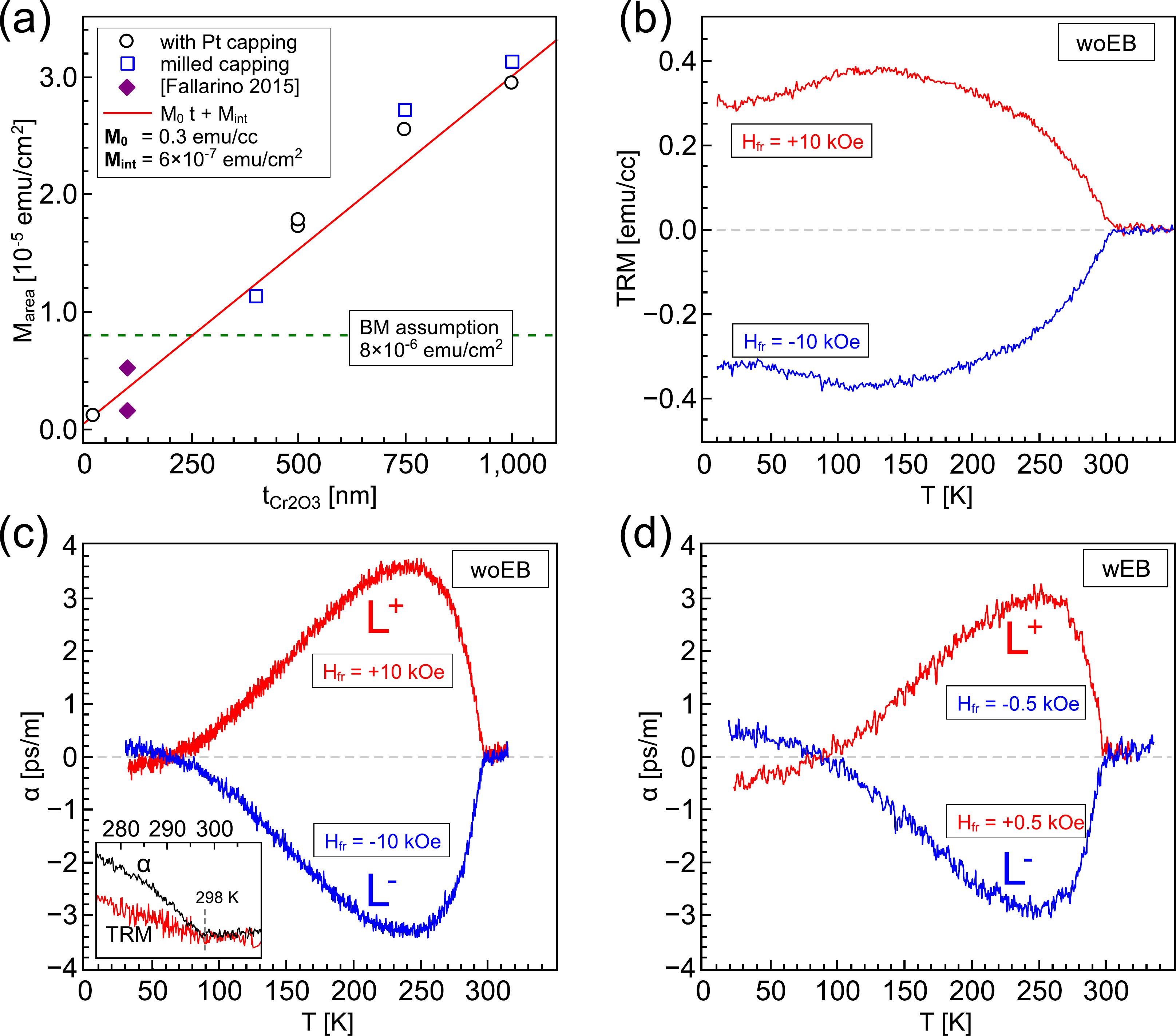}
\end{figure}

The temperature dependence of the thermoremnant magnetization (TRM) at zero field after a magnetic-field cooling was measured for various thicknesses of \cro with Pt capping. If the origin of $M_0$ is the fully polarized boundary magnetization at the \cro surface, then the areal density of magnetization should be independent of thickness. However, the magnetization areal density showed a linear dependence on thickness, with a slope of 0.3 emu/cc and a negligible value of the interface component [Fig.~\ref{fig:coupling}(a)]. After the removal of capping by Ar$^+$ ion milling, no significant change of magnetization was found [blue squares in Fig.~\ref{fig:coupling}(a)]. The estimated fully polarized surface should have an areal magnetization of $8\times 10^{-6}$ emu/cm$^2$ that is thickness-independent [dashed green line in Fig.~\ref{fig:coupling}(a)] \cite{fallarino_2015}. TRM switched in the same direction as $H_\mathrm{fr}$ [Fig.~\ref{fig:coupling}(b)]. The temperature dependence of $\alpha$ also switched with the direction of $H_\mathrm{fr}$ [Fig.~\ref{fig:coupling}(c)]. At 240 K, $\alpha$ was positive (negative) for a positive (negative) $H_\mathrm{fr}$. Therefore, the bulk $M_0$ is parallel to $l$ [the left-hand-side of Fig.~\ref{fig:schem}(b)]. The AFM domains are coupled to $H_\mathrm{fr}$ through $M_0$ [See appendix \ref{sec:appx_M_l}, and the Zeeman energy prefers $l \parallel H_\mathrm{fr}$. Furthermore, both the spontaneous magnetization and the AFM order became zero at the same transition temperature of 298 K. The inset of Fig.~\ref{fig:coupling}(c) shows TRM and $\alpha$ measured for the same cut of the sample woEB. 

No other phases of chromium oxide were found by X-ray diffraction analysis \cite{pati_2015,borisov_2016,shimomura_2016}, and the oxidation state of Cr ions was close to $+$3 as confirmed by X-ray photoelectron spectroscopy and X-ray absorption near the edge structure (data not shown). All these mean that the observed spontaneous volume magnetization originates from the ordering of Cr spins in \cro. As a speculation on the origin, we should consider the magnetic moments on each sublattice. The magnetic moments of Cr ions are reduced from an expected maximum of 3 $\mu_B$, due to the charge back-transferred from oxygen 2\emph{p} orbitals into Cr 3\emph{d} orbitals \cite{brown_2002}. In the sputtered films, the unit cell is distorted and strained \cite{pati_2015,borisov_2016}. It is likely that the atom coordinates of fabricated films are changed from bulk values. The different bond lengths and angles of Cr-O can give a different magnetic moment per sublattice. This would result in a ferrimagnetic-type order, represented by different arrow sizes in Fig.~\ref{fig:schem}(b). The corresponding change in Cr magnetic moments needs to be 2$ \times 10^{-3}\mu_B$ per ion to account for $M_0$. The relative orientation of $M_0$ as parallel or antiparallel to $l$ depends on the details of which sublattice is dominant.

The relation between the direction of Co magnetization and the \cro AFM vector is antiparallel based on two considerations. First, it was reported that the Cr and Co spins are antiferromagnetically exchange-coupled \cite{shiratsuchi_2015}. The second consideration is that corundum-type crystals prefer to terminate in the bottom half of the buckled metal-ion layer for most fabrication conditions \cite{wang_2000}. This termination corresponds to the bottom spin sublattice of Fig.~\ref{fig:schem}(a). Therefore, the direction of Cr$_2$O$_3$ surface spin is parallel to $l$. The combination of the previous two observations indicates a preference of an $L^-$ domain by the exchange coupling to Co spins for $H_\mathrm{fr} > 0$ [the right-hand-side of Fig.~\ref{fig:schem}(b)]. Figure \ref{fig:coupling}(d) shows the measurement of the direction of AFM domains in relation to $H_\mathrm{fr}$ for the sample wEB. The negative (positive) $\alpha_\mathrm{peak}$ for a positive (negative) $H_\mathrm{fr}$ indicates that $L^-$ ($L^+$) domains are stabilized by coupling to Co spins [Fig.~\ref{fig:coupling}(d)]. Hence, the exchange coupling prefers an antiparallel relation between $l$ and $H_\mathrm{fr}$. More confirmation on the coupling sign between Co and \cro domains is presented in appendix \ref{sec:appx_M_l}. The sign of coupling did not change by changing the material of metal spacer or by removing the spacer.

\begin{figure}
	\caption{(a) The temperature dependence of $\alpha$ in the sample woEB, normalized by the maximum peak value for a cooling under the simultaneous application of a magnetic field $H_\mathrm{fr}=+10$ kOe and a varying electric field $E_\mathrm{fr}$. An additional negative $E_\mathrm{fr}$ was needed to switch the Cr$_2$O$_3$ domains. (b) A comparison of the dependence of the average domain state $\langle L \rangle$ on $E_\mathrm{fr}$ between the samples woEB and wEB. The sample woEB shows a constant threshold electric field $E_\mathrm{th}$, irrespective of $H_\mathrm{fr}$. The sample wEB has a shift of $E_\mathrm{th}$ with changing $H_\mathrm{fr}$. At $H_\mathrm{fr}\approx 2.6$ kOe, the switching curve has a zero $E_\mathrm{th}$. (c) Plots of $E_\mathrm{th}$--$1/H_\mathrm{fr}$ with fits to the model outlined in Eq.~\ref{eq:Eth}. A single-crystal slab of Cr$_2$O$_3$ has a zero $E_\mathrm{th}$ regardless of $1/H_\mathrm{fr}$ [blue diamonds]. The samples wEB and woEB have similar negative \emph{y}-intercepts, due to the presence of $M_0$ [black circles and red squares, respectively]. For the sample wEB, a slope results from exchange coupling to Co, and a zero $E_\mathrm{th}$ is found at $H_{E0}$. (d) The dependence of $\lambda_E$ on $1/H_\mathrm{fr}$ is linear with a slope $\mathit{EH}_T$ that is determined by $\alpha V$. For the deposited films, a nonzero $\lambda_E$ is still present even in the vanishing $1/H_\mathrm{fr}$ limit ($C$).}
	\label{fig:MEFC}
	\includegraphics[width=0.6\textwidth]{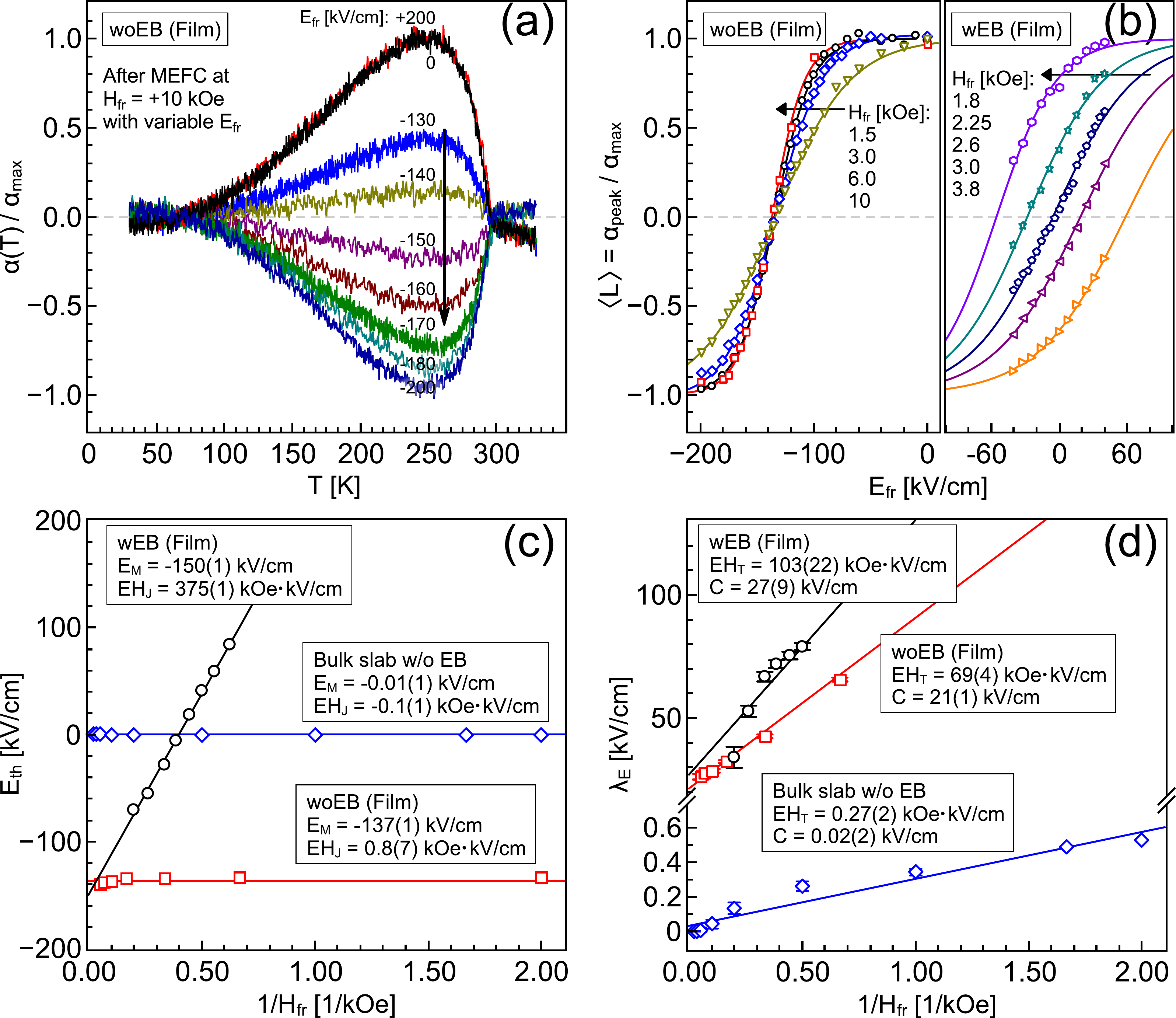}
\end{figure}

Between the two AFM domain states $L^\pm$, the sign of $\alpha$ changes but with the same magnitude. Therefore, the normalized $\alpha$ by the saturation value gives the average domain state as $\langle L \rangle = \alpha_\mathrm{peak}/\alpha_\mathrm{max}$. Figure \ref{fig:MEFC}(a) shows the temperature dependence of $\alpha$ for the sample woEB after MEFC under $H_\mathrm{fr}$ = 10 kOe and a variable $E_\mathrm{fr}$. For the whole range of temperatures, $\alpha$ changes linearly between two opposite maxima, and the dependence of normalized $\alpha$ on cooling fields is temperature independent. Thus, it is suitable to estimate $\langle L \rangle$ from measuring only the normalized peak value at 240 K. The measurement was performed after MEFC at $H_\mathrm{fr}$ and $E_\mathrm{fr}$ fields from 320 K, which is higher than $T_N$, down to the detection temperature of 240 K. $\alpha$ is presented in ps/m for consistency with other reports. The peak-to-peak noise level was $< 5 \times 10^{-9}$ emu, and the measurement was averaged over a 3-min span.

The measurements of $\langle L \rangle$--$E_\mathrm{fr}$ curves were used to find $E_\mathrm{th}$ from fittings to Eq.~\ref{eq:Ftanh} for a varying $H_\mathrm{fr}$ [Fig.~\ref{fig:MEFC}(b)]. The data points were taken at a random sequence, and no effect of points ordering was found. The left-hand-side of Fig.~\ref{fig:MEFC}(b) shows $\langle L \rangle$--$E_\mathrm{fr}$ dependence of the sample woEB. A gradual switching between the positive and negative maxima was found, with a tendency closely resembled by Eq.~\ref{eq:Ftanh}. $E_\mathrm{th}$ is at a constant $-137$ kV/cm, not dependent on $H_\mathrm{fr}$. During cooling in a positive $H_\mathrm{fr}$, an additional negative electric field is needed so that the negative magnetoelectric energy favoring the $L^-$ state overcomes the Zeeman energy of $M_0$ favoring the $L^+$ state. As the magnetic field is increased, both Zeeman and magnetoelectric energies increase by the same amount. Therefore, $E_\mathrm{th}$ is not dependent on $H_\mathrm{fr}$ [the red squares in Fig.~\ref{fig:MEFC}(c)]. In comparison, a bare 0.5-mm slab of a single-crystal Cr$_2$O$_3$ had a zero $E_\mathrm{th}$ irrespective of $1/H_\mathrm{fr}$, due to the absence of $M_0$ and $J_K$ [the blue diamonds in Fig.~\ref{fig:MEFC}(c)]. In the sample wEB, the $\langle L \rangle$--$E_\mathrm{fr}$ dependence shifts with $H_\mathrm{fr}$ due to the presence of $J_K$ [Fig.~\ref{fig:MEFC}(b)]. At $H_\mathrm{fr} \approx$ 2.6 kOe, the $\langle L \rangle$--$E_\mathrm{fr}$ curve is symmetric around a zero $E_\mathrm{fr}$. This demonstrates that low-voltage switching at a finite magnetic field is possible. The shift of $E_\mathrm{th}$ is linear with $1/H_\mathrm{fr}$ [the black circles in Fig.~\ref{fig:MEFC}(c)], and the \emph{y}-intercept is in agreement with the sample woEB. From Eq.~\ref{eq:Eth}, the \emph{x}-intercept corresponds to $H_{E0} = J_K /( M_0 t)$. An estimation from the low-temperature values of $J_K$ and $M_0$ gives $H_{E0}\approx 5.3$ kOe. Considering the approximations in the macrospin model, the experimental value of $H_{E0} = 2.6$ kOe has a reasonable agreement. The presence of $M_0$ can be used to estimate the effective $\alpha$ at $T_N$ during MEFC from $E_M = -M_0/\alpha$, which is estimated at 27 ps/m. 

Next, we turn to the smooth switching around $E_\mathrm{th}$, represented by $\lambda_E$. $\lambda_E$ characterizes the required $\lvert E_\mathrm{fr} - E_\mathrm{th} \rvert$ for a saturation of the switching probability. It is linearly proportional to $1/H_\mathrm{fr}$, where the proportionality constant is the thermal fluctuation energy translated into ME energy $\mathit{EH}_T = 2 k_B T_N / \alpha V$. Figure \ref{fig:MEFC}(d) shows the $\lambda_E$--$1/H_\mathrm{fr}$ plot for both of the samples wEB and woEB [black circles and red squares, respectively]. The slope of $\lambda_E$--$1/H_\mathrm{fr}$ line is finite, corroborating the finite value of $\alpha V$. However, a nonzero $\lambda_E$ is found when $1/H_\mathrm{fr}$ approaches zero. For a higher $H_\mathrm{fr}$, the switching probability should saturate for a smaller $\lvert E_\mathrm{fr} - E_\mathrm{th} \rvert$ due to the larger ME energy. However, short-ranged paramagnetic spin waves are reported to be excited at the phase-transition temperature  or at a higher temperature \cite{hertz_1977,moriya_1982}. Spin-wave fluctuations with energy that is dependent on $H_\mathrm{fr} M_0$ can explain why a nonzero $\lvert E_\mathrm{fr} - E_\mathrm{th} \rvert$ is still needed at a high $H_\mathrm{fr}$ in the sputtered films. Due to the large $V$ of the single-crystal slab and the absence of $M_0$, a line with a much smaller slope and no intercept was found [blue diamonds in Fig.~\ref{fig:MEFC}(d)].

\section{Summary}
In this work, we described the magnetoelectric switching of the antiferromagnetic domains of Cr$_2$O$_3$ films by using a phenomenological model, and compared it with direct measurements on the average domain state after cooling in magnetic and electric fields. The model is based on the observation of a competition between the Zeeman energy of a spontaneous volume magnetization of Cr$_2$O$_3$ sputtered films, and the exchange-coupling energy with a proximate ferromagnetic layer. We found that the volume magnetization of Cr$_2$O$_3$ films caused an opposite shift in the required electric field for switching compared to the shift caused by the exchange coupling. At the point of balance, low values of the writing electric field were achieved that are symmetric around a zero threshold value. We expect that such a development can pave the way for controlling both the writing energy and the exchange bias in heterostructure-based magnetoelectric and multiferroic information media.
\begin{acknowledgments}
This work was partly funded by the ImPACT Program of the Council for Science, Technology and Innovation (Cabinet Office, Japan Government).
\end{acknowledgments}

\appendix
\section{Coupling sign of Co and Cr$_2$O$_3$ domains and the linearity of magnetoelectric susceptibility $\alpha$}\label{sec:appx_alpha}
\begin{figure}
	\caption{}
	\label{fig:Mac}
	\includegraphics[width=0.4\textwidth]{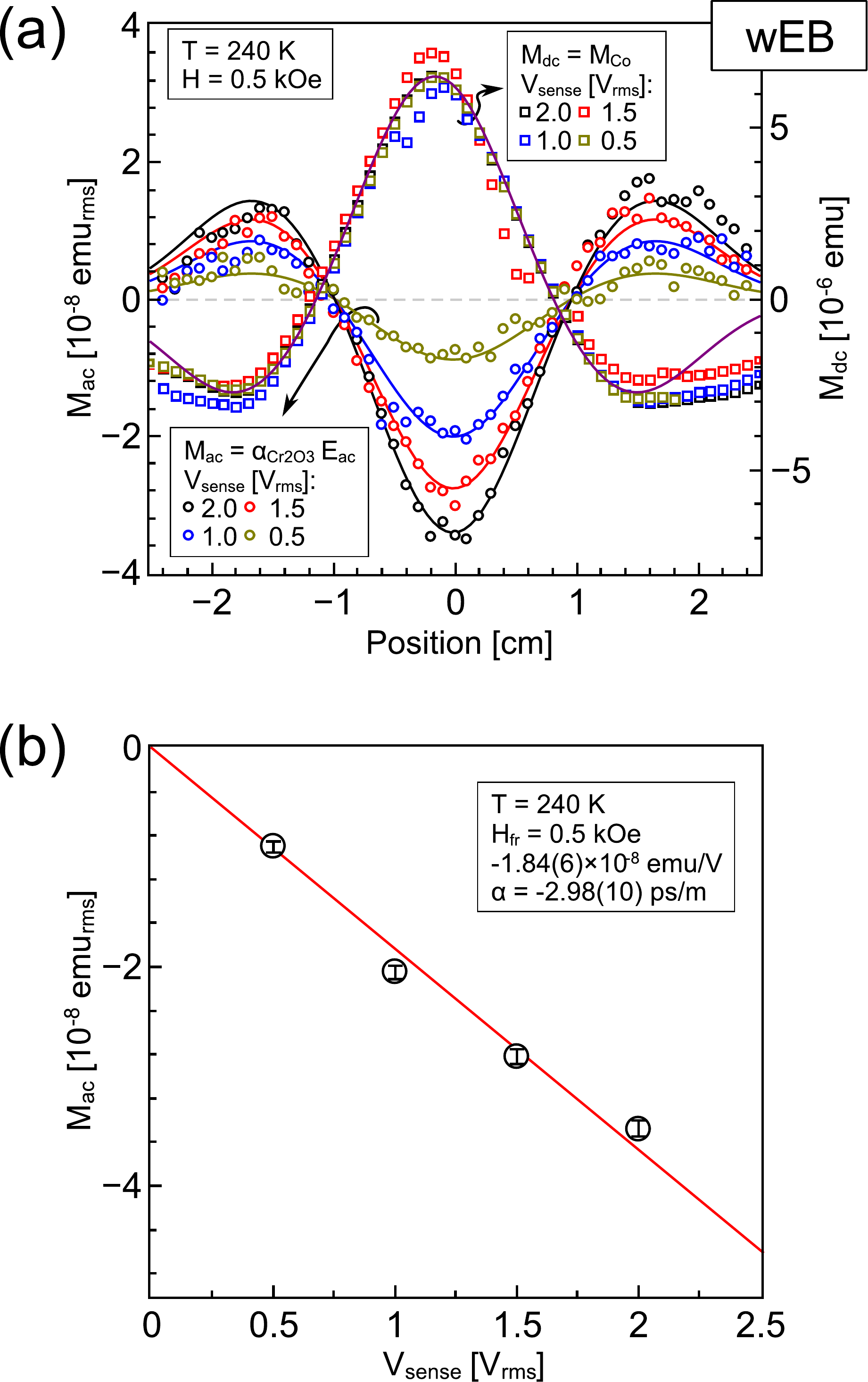}
\end{figure}

We measured the ME effect in the sample wEB after magnetic field cooling. The simultaneous measurement of the \emph{dc} and \emph{ac} responses of the SQUID magnetometer along the detection coils is shown in Fig.~\ref{fig:Mac}(a). The measurement was at 240 K with $+$500 Oe applied along the detection coils after cooling under $+$500 Oe from 320 K. The shape of the response was typical for a second-order gradiometer. The $M_\mathrm{dc}$ is mainly coming from Co due to its high saturation magnetization. The direction of Co magnetization is parallel to $H_\mathrm{fr}$ as expected. The $M_\mathrm{ac} = \alpha_\mathrm{Cr2O3} E_\mathrm{ac}$ is from the ME effect of \cro layer. The sign of $M_\mathrm{ac}$ is negative corroborating an $L^-$ domain state. The $L^-$ domains are stabilized by the antiferromagnetic coupling of \cro surface ions to Co spins.

The amplitude of $M_\mathrm{ac}$ increased by increasing the sensing voltage $V_\mathrm{sense}$, whereas $M_\mathrm{dc}$ remained unchanged as expected from the absence of an ME effect in Co. The dependence of $M_\mathrm{ac}$ on $V_\mathrm{sense}$ showed a linear response indicating that the magnetoelectric effect is indeed linear [Fig.~\ref{fig:Mac}(b)].

\section{Characteristics of the sample with exchange-bias}\label{sec:appx_Hex}
\begin{figure}
	\caption{}
	\label{fig:Hex_Hfr}
	\includegraphics[width=0.4\textwidth]{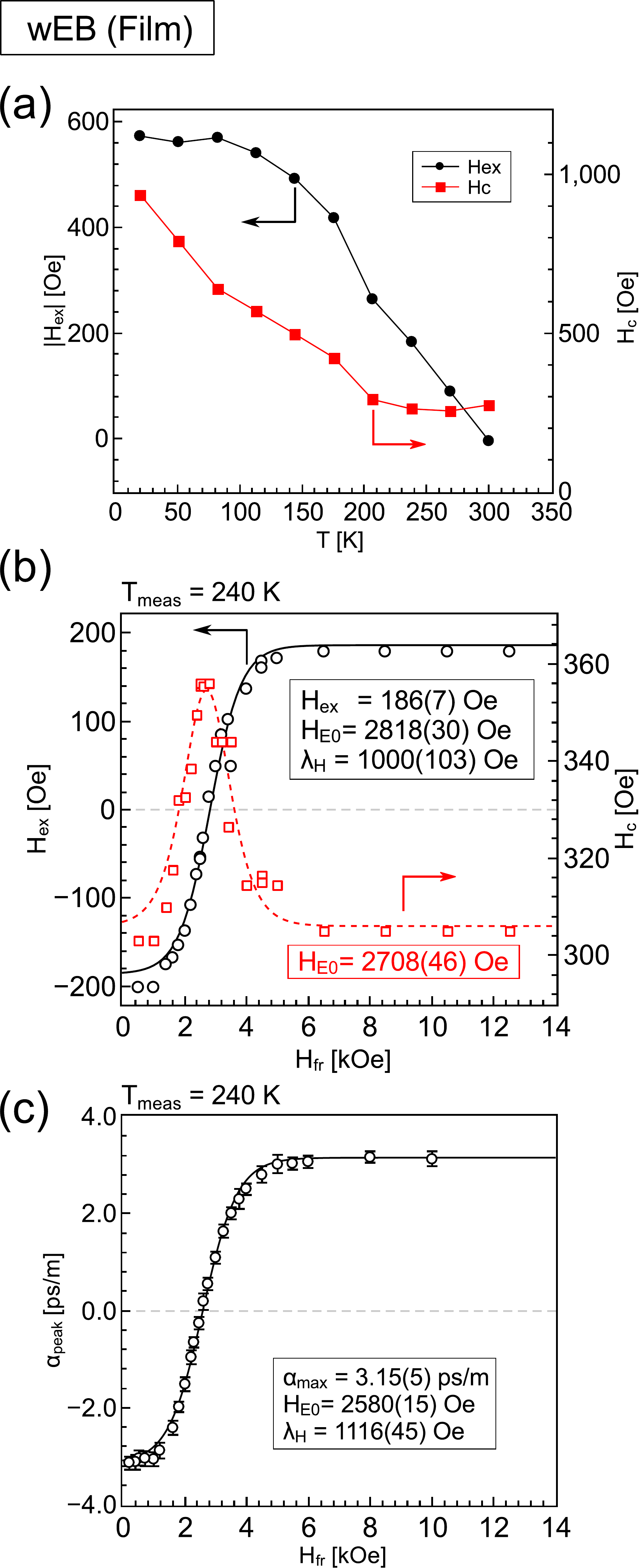}
\end{figure}

The exchange bias $H_\mathrm{ex}$ was measured from the shift in the magnetization hysteresis loop of the Co layer using a reciprocating sample option of the SQUID magnetometer. The exchange-bias data were acquired from a sample that was prepared under the same fabrication conditions as the sample wEB. Figure \ref{fig:Hex_Hfr}(a) shows the temperature dependence of exchange-bias field $H_\mathrm{ex}$ and coercivity $H_c$. The low-temperature limit of $H_\mathrm{ex}$ was 570 Oe, corresponding to $J_K = 8 \times 10^{-3} \mathrm{erg/cm}^2$.
 
In the main text of the paper, $\alpha_\mathrm{peak}$ was used as the main indication of the average domain state $\langle L \rangle$. The measurement of $\alpha$ is related to the bulk domain state. An indirect measurement is the exchange bias on a ferromangetic layer, which is representative of the domain state at interface. Here we compare both methods. 

Figures \ref{fig:Hex_Hfr}(b) and \ref{fig:Hex_Hfr}(c) show the switching of $\langle L \rangle$ after magnetic-field-cooling in $H_\mathrm{fr}$ with a zero electric field. Figure \ref{fig:Hex_Hfr}(b) shows the switching measured by $H_\mathrm{ex}$ at 240 K. In the intermediate switching region 2--4 kOe, the \cro film is composed of a large number of $L^\pm$ domains, and the exchange-bias on the Co layer is weaker that the exchange stiffness of Co. Therefore, the normalized $H_\mathrm{ex}$ can be treated as the total average $\langle L \rangle$. The solution to Eq.~4a of the main text in the absence of electric field during cooling gives:

\begin{equation}\label{eq:Ltanh}
\langle L \rangle = \tanh \left( \frac{H_\mathrm{fr}-H_{E0}}{\lambda_H} \right) \,,
\end{equation}
where:

\begin{align}
H_{E0} =& \frac{J_K}{M_0 t}\, , \\
\mathrm{and}\quad 
\lambda_H =& \frac{2 k_B T_N}{M_0 V} \, .
\end{align}

The fittings to Eq.~\ref{eq:Ltanh} are shown as solid lines, where the fitting values are inside the boxes. The two measurement methods of Figs.~\ref{fig:Hex_Hfr}(b) and \ref{fig:Hex_Hfr}(c) are in a good agreement. The coercivity of Co $H_c$ has a maximum at $H_{E0}$, where an equal number of $L^\pm$ domains are present [Fig.~\ref{fig:Hex_Hfr}(b), with a red dashed line as an eye-guide].

\section{Coupling of the antiferromagnetic domains and the spontaneous magnetization of sputtered films}\label{sec:appx_M_l}
\begin{figure*}
	\caption{}
	\label{fig:Hrev}
	\includegraphics[width=0.8\textwidth]{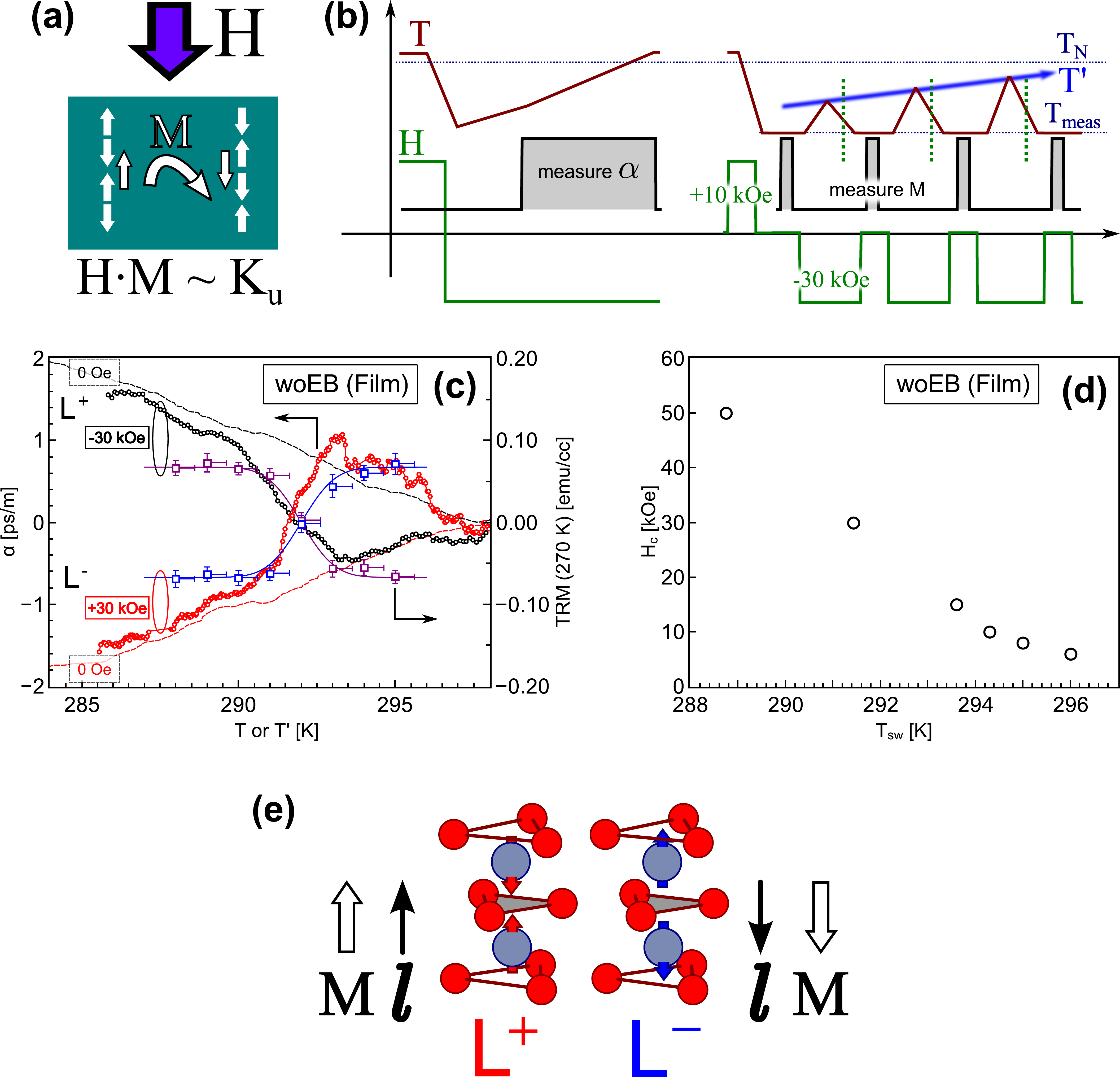}
\end{figure*}

In the main text, it is shown that the spontaneous magnetization $M_0$ and the AFM vector point in the same direction. To find how strong this spontaneous magnetization is coupled to AFM vector, we checked the effect of an external magnetic field on the switching of AFM domains in sample woEB [Fig.~\ref{fig:Hrev}(a)]. After MFC at a certain $H_\mathrm{fr}$, a field in the opposite direction was applied, then $\alpha$ and $M$ were measured while the temperature was increased at a heating rate of $+$0.5 K/min [Fig.~\ref{fig:Hrev}(b)]. After MFC at $+$10 kOe, a field of $-$30 kOe was set and $\alpha$-T dependence was measured. $\alpha$ changed from positive to negative values during the temperature increase [black circles in Fig.~\ref{fig:Hrev}(c)]. Comparing with the measurements at zero field [dashed lines in Fig.~\ref{fig:Hrev}(c)], we can say that the change corresponds to switching from $L^+$ to $L^-$. The same opposite results were found for $-$10-kOe MFC and subsequent $+$30-kOe measurement [red circles in Fig.~\ref{fig:Hrev}(c)]. 

The $M$--$T$ dependence could not be measured directly due to the large diamagnetic background of the substrate. We resorted to the procedure of Ref.~\onlinecite{fallarino_2014}, by applying the opposing magnetic field up to a certain temperature $T^\prime$, then cooling to 270 K and measuring TRM at a zero field [Fig.~\ref{fig:Hrev}(b)]. The temperature overshoot at $T^\prime$ was minimized to $<0.6$ K. The switching of $M$ against $T^\prime$ was at the same temperature as $\alpha$ [squares in Fig.~\ref{fig:Hrev}(c)]. This confirms that $l$ switches together with $M$.

The switching temperature $T_\mathrm{sw}$ corresponds to the coercive field $H_c$ at that temperature, and the dependence shows a large temperature variation similar to Ref.~\onlinecite{fallarino_2014} [Fig.~\ref{fig:Hrev}(d)]. This large variation is likely due to the large decrease of uniaxial magnetocrytalline anisotropy near $T_N$ \cite{foner_1963}. As the thermal energy increases and anisotropy decreases, the thermally activated switching occurs when anisotropy $K_u$ becomes on same order of magnitude as the Zeeman energy of Cr$_2$O$_3$ magnetization $H \cdot M_0$.

We conclude by restating that $l$ is parallel to $M_0$ [Fig.~\ref{fig:Hrev}(e)].

\bibliographystyle{apsrev4-1}
\bibliography{../../ZotPapers}
\clearpage
\newpage
\begin{figure}
	\caption{(a) Definitions of AFM domains of Cr$_2$O$_3$ with respect to the directions of applied fields. The AFM-staggered magnetization vector $l$ is defined as parallel to the spin sublattice beneath the smaller oxygen triangle (grayed triangle area). An angle $\theta = 0$($\pi$) between $l$ and the positive \emph{z}-axis, corresponds to the $L^+$($L^-$) domain. An $L^+$($L^-$) domain results from a parallel (antiparallel) magnetoelectric field cooling, and has a positive (negative) value of magnetoelectric susceptibility $\alpha$. (b) For a positive magnetic field, the Zeeman energy of Cr$_2$O$_3$ spontaneous volume magnetization $M_0$, which is parallel to $l$, prefers an $L^+$ domain [left]. On the other hand, the interfacial exchange coupling with the Co ferromagnetic layer favors an $L^-$ domain [right].}
	\label{fig:schem}
	\includegraphics[width=0.45\textwidth]{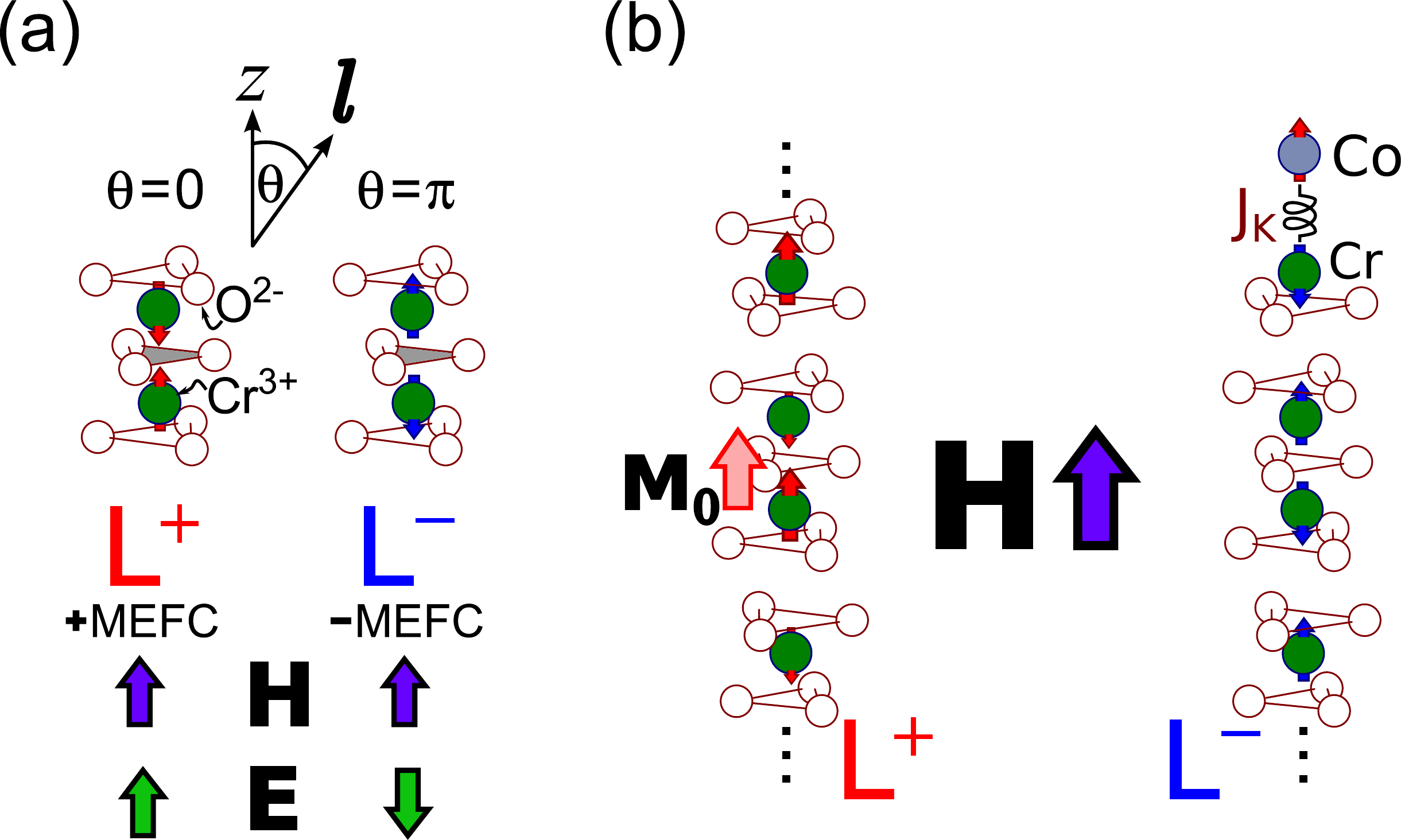}
\end{figure}

\end{document}